# Polarizabilities of microparticles in relativistic field theory

N.V. Maksimenko, O.M. Deruzhkova, S.A. Lukashevich

*Gomel State University, Sovetskay street, 102, Gomel, Belarus, 246019;*

*E-mail: lukashevich@gsu.by*

(August 21, 2021)

## Abstract

The article   determine a relativistic tensor of the second rank containing the vectors of the electric $\vec{P}$  and magnetic $\vec{M}$ polarization of the medium, based on the Maxwell equations and the determination of the charge density and polarization current of a structural microparticle. Using this tensor and the electromagnetic field tensor, the relativistic lagrangian of the interaction an electromagnetic field with a structural microparticle is obtained, taking into account the polarization of its structural elements. Based on the induced moments, the relativistic lagrangian of the interaction an electromagnetic field with a structural microparticle is constructed with considering the electric and magnetic polarizabilities.

One of the effective theoretical methods for studying electrodynamic processes is using of lagrangians obtained within the framework of relativistic field approaches and agreed with low-energy theorems [1].

The construction of effective relativistic-invariant lagrangians allows to establish not only the physical interpretation of the electromagnetic characteristics of microparticles, but also to obtain information about the mechanisms of the interaction an electromagnetic field with such kind of particles.

The most important electromagnetic characteristics of microparticles are their polarizabilities, the value of which is due by the structural and quantum properties of these microparticles [2].

Recently, many electrodynamic processes have been used, on the basis of which is realizing data on the polarizabilities of such microparticles as hadrons, nuclei, etc.

In this connection, the problem arises of how to consistently take into account the contribution of polarizabilities to the amplitudes and cross-sections of electrodynamic processes on microparticles in a covariant way. Such a problem can be solved by constructing a field relativistic formalism of the interaction an electromagnetic field with microparticles, taking into account their polarizabilities.

In [3], the effective relativistic lagrangian of the interaction an electromagnetic field with microparticles containing the electric and magnetic dipole moments was constructed.

Here, in this paper, using the methods of articles [3,4,5] the relativistic lagrangian is obtained taking into account the electric and magnetic polarizabilities based on induced dipole moments. Some consequences from this lagrangian are established.

Consider the polarizabilities of microparticles in relativistic field theory. By definition, the electromagnetic field tensor $F^{\mu\nu}$ is expressed in terms of the four-dimensional potential an electromagnetic field

$$A^\mu \{ A^0 = \varphi, A^i \}$$

as follows

$$F^{\mu\nu} = \partial^\mu A^\nu - \partial^\nu A^\mu, \tag{1.1}$$

where

$$\partial^\mu \{ \partial^0 = \partial_t, -\vec{\nabla} \},$$
$$\partial_\mu \{ \partial_0 = \partial_t, \vec{\nabla} \}.$$

Thus, the explicit form of the tensor $F^{\mu\nu}$ (1.1) is represented as a matrix

$$F^{\mu\nu} = \begin{pmatrix} 0 & -E^1 & -E^2 & -E^3 \\ E^1 & 0 & -B^3 & B^2 \\ E^2 & B^3 & 0 & -B^1 \\ E^3 & B^2 & B^1 & 0 \end{pmatrix},$$

(1.2)

where $E^i$ — are the components of the vector of electric field strength, $B^i$ — are the components of the magnetic induction vector. Using the metric tensor definition

$$g_{\mu\nu} = \begin{pmatrix} 1 & 0 & 0 & 0 \\ 0 & -1 & 0 & 0 \\ 0 & 0 & -1 & 0 \\ 0 & 0 & 0 & -1 \end{pmatrix},$$

(1.3)

we obtain the following definition of the electromagnetic field tensor

$$F_{\mu\nu} = g_{\mu\rho} g_{\nu\sigma} F^{\rho\sigma},$$

which in the form of a matrix has the form:

$$F_{\mu\nu} = \begin{pmatrix} 0 & E^1 & E^2 & E^3 \\ -E^1 & 0 & -B^3 & B^2 \\ -E^2 & B^3 & 0 & -B^1 \\ -E^3 & -B^2 & B^1 & 0 \end{pmatrix}.$$

(1.4)

Using the Levi-Civita tensor $\varepsilon^{\mu\nu\rho\sigma}$ (in this paper we use the condition $\varepsilon^{0123} = +1$, $\varepsilon_{0123} = -1$), we determine the dual electromagnetic tensors

$$\tilde{F}^{\mu\nu} = \frac{1}{2} \varepsilon^{\mu\nu\rho\sigma} F_{\rho\sigma} = \begin{pmatrix} 0 & -B^1 & -B^2 & -B^3 \\ B^1 & 0 & E^3 & -E^2 \\ B^2 & -E^3 & 0 & E^1 \\ B^3 & E^2 & -E^1 & 0 \end{pmatrix},$$

(1.5)

and

$$\tilde{F}_{\mu\nu} = \frac{1}{2} \varepsilon_{\mu\nu\rho\sigma} F^{\rho\sigma} = \begin{pmatrix} 0 & B^1 & B^2 & B^3 \\ -B^1 & 0 & E^3 & -E^2 \\ -B^2 & -E^3 & 0 & E^1 \\ -B^3 & E^2 & -E^1 & 0 \end{pmatrix}.$$

(1.6)

From the definitions (1.2−1.6) follows the expression for the lagrangian of the electromagnetic field

$$\mathcal{L} = -\frac{1}{4} F_{\mu\nu} F^{\mu\nu} = -\frac{1}{2} (\vec{B}^2 - \vec{E}^2),$$

as well as the ratio

$$\tilde{F}_{\rho\varkappa}\tilde{F}^{\rho\sigma} = F_{\rho\varkappa}F^{\rho\sigma} - \frac{1}{2}\delta_\varkappa^\sigma F^2,$$

where $F^2 = F_{\mu\nu}F^{\mu\nu}$.

Based on the matrix representation of the electromagnetic tensors (1.2−1.6) and the four-dimensional vector of the particle (the system of units $\hbar = c = 1$ is used)

$$u^\mu\{u^0, \vec{u}\}, \qquad u^2 = (u^0)^2 - \vec{u}^2 = 1,$$

where $u^0 = \gamma = \frac{1}{\sqrt{1-\beta^2}}$, $\vec{u} = \gamma\vec{v}$, $\vec{v}$ − the speed of the particle, $\beta^2 = \vec{v}^2$, define following four-dimensional vectors [6]

$$e^\mu = F^{\mu\nu}u_\nu, \ e_\mu = F_{\mu\nu}u^\nu, \ b^\mu = \tilde{F}^{\mu\nu}u_\nu, \ b_\mu = \tilde{F}_{\mu\nu}u^\nu. \qquad (1.7)$$

These vectors are expressed in terms of the vectors $\vec{E}$, $\vec{B}$ and the particle velocity as follows:

$$e^\mu\{\vec{u}\vec{E}, (u^0\vec{E} + [\vec{u}\vec{B}])\}, \qquad (1.8)$$

$$e_\mu\{\vec{u}\vec{E}, -(u^0\vec{E} + [\vec{u}\vec{B}])\}, \qquad (1.9)$$

$$b^\mu\{\vec{u}\vec{B}, (u^0\vec{B} - [\vec{u}\vec{E}])\}, \qquad (1.10)$$

$$b_\mu\{\vec{u}\vec{B}, (-u^0\vec{B} + [\vec{u}\vec{E}])\}. \qquad (1.11)$$

In its turn the tensors of an electromagnetic field (1.2−1.6) can be determined using the vectors (1.7−1.10) and the vector $u^\mu$. So, for example,

$$F^{\mu\nu} = e^\mu v^\nu - e^\nu u^\mu + \varepsilon^{\mu\nu\rho\sigma}u_\rho b_\sigma, \qquad (1.12)$$

We now turn to the relativistic field description of the interaction  an electromagnetic field with particles whose electric and magnetic polarizabilities are nonzero.

In non-relativistic electrodynamics, the electric $\alpha_E$ and magnetic $\beta_M$ polarizabilities are the quantities by which the proportionality between the applied electric and magnetic fields and the induced dipole moments is established  [8]:

$$\vec{d} = 4\pi\alpha_E\vec{E}, \qquad (1.13)$$

$$\vec{m} = 4\pi\beta_M\vec{H}. \qquad (1.14)$$

Therefore, polarizabilities have a fundamental property and are related to the structure of microparticles and the mechanism of interaction an electromagnetic field with these microsystems.

The interaction energy of an applied electromagnetic field with the microparticle has the form:

$$u = -\frac{4\pi}{2}\alpha_E \vec{E}^2 - \frac{4\pi}{2}\beta_{\text{M}}\vec{H}^2.$$

$$(1.15)$$

If a microsystem consists of $N$ particles per unit volume, then the dielectric permittivity is related to the polarizabilities by the relations [8]

$$\varepsilon = 1 + N4\pi\alpha_E, \ \mu = 1 + N4\pi\beta_M. \qquad (1.16)$$

It can be seen from these relations that the polarizabilities $\alpha_E$ and $\beta_M$ have the dimension cm$^3$.

In turn, the electric and magnetic polarizations of the macrosystem are determined in terms of polarizabilities as follows:

$$\vec{P} = N4\pi\alpha_E\vec{E}, \quad \vec{M} = N4\pi\beta_M\vec{H}. \qquad (1.17)$$

To go to the relativistic view of the relation (1.17) let's use the Maxwell equations taking into account the polarization of the medium [7]

$$rot\vec{B} = \frac{\partial\vec{E}}{\partial t} + \vec{j} + \vec{j_P}, \qquad (1.18)$$

$$div\vec{E} = \rho + \rho_{\text{P}}, \qquad (1.19)$$

$$div\vec{B} = 0, \qquad (1.20)$$

$$rot\vec{E} = -\frac{\partial\vec{B}}{\partial t}. \qquad (1.21)$$

In equations (1.18) and (1.19), $\rho$ and $\vec{j}$ are the densities of free charges and currents, $\rho_P$ and $\vec{j}_P$ are the densities of charges and currents which are caused by polarization.

The value $\rho_{\text{P}}$ and $\vec{j}_P$ are expressed in terms of $\vec{P}$ and $\vec{M}$ by following

$$\rho_{\text{P}} = -(\vec{\nabla}\vec{P}), \qquad (1.22)$$

$$\vec{j}_P = \frac{\partial\vec{P}}{\partial t} + rot\vec{M}. \qquad (1.23)$$

The relations (1.22−1.23) can be represented in a covariant form

$$j_P^\nu = \partial_\mu M^{\mu\nu}, \qquad (1.24)$$

where $\mu$ takes the values 0,1,2,3.

In the definition of a four-dimensional current due to polarization, the tensor $M^{\mu\nu}$ has the form:

$$M^{\mu\nu} = \begin{pmatrix} 0 & P^1 & P^2 & P^3 \\ -P^1 & 0 & -M^3 & M^2 \\ -P^2 & M^3 & 0 & -M^1 \\ -P^3 & -M^2 & M^1 & 0 \end{pmatrix}. \tag{1.25}$$

If we proceed to the three-dimensional recording of the ratio (1.24) using (1.25), we get

$$j_P^{(0)} = \rho_P = -(\vec{\nabla}\vec{P}), \ \vec{j}_P = \partial_t \vec{P} + [\vec{\nabla}\vec{M}].$$

Using the $g_{\mu\nu}$ metric tensor and the Levi-Civita tensor, we can determine the matrix representation of the tensors $M_{\mu\nu}$, $\widetilde{M}^{\mu\nu}$, $\widetilde{M}_{\mu\nu}$. The result of the calculations is represented by the relations:

$$M_{\mu\nu} = \begin{pmatrix} 0 & -P^1 & -P^2 & -P^3 \\ P^1 & 0 & -M^3 & M^2 \\ P^2 & M^3 & 0 & -M^1 \\ P^3 & -M^2 & M^1 & 0 \end{pmatrix},$$

$$\widetilde{M}^{\mu\nu} = \begin{pmatrix} 0 & -M^1 & -M^2 & -M^3 \\ M^1 & 0 & -P^3 & P^2 \\ M^2 & P^3 & 0 & -P^1 \\ M^3 & -P^2 & M^1 & 0 \end{pmatrix},$$

$$\widetilde{M}_{\mu\nu} = \begin{pmatrix} 0 & M^1 & M^2 & M^3 \\ -M^1 & 0 & -P^3 & P^2 \\ -M^2 & P^3 & 0 & -P^1 \\ -M^3 & -P^2 & P^1 & 0 \end{pmatrix} \tag{1.26}$$

Based on the tensors $M^{\mu\nu}, M_{\mu\nu}, \widetilde{M}^{\mu\nu}$ and $\widetilde{M}_{\mu\nu}$ we define the four-dimensional vectors

$$d^\mu = M^{\mu\nu}u_\nu, \ d_\mu = M_{\mu\nu}u^\mu, \ m^\mu = \widetilde{M}^{\mu\nu}u_\nu, m_\mu = \widetilde{M}_{\mu\nu}u^\nu. \tag{1.27}$$

Similarly to the definition of the electromagnetic field tensor $F^{\mu\nu}$ using the vectors $e^\mu, e_\mu, b^\mu$ и $b_\mu$ (1.12), the tensor $M^{\mu\nu}$ can be represented

$$M^{\mu\nu} = d^\mu u^\nu - d^\nu u^\mu + \varepsilon^{\mu\nu\rho\sigma}u_\rho m_\sigma. \tag{1.28}$$

From relations (1.25) and (1.26), it follows that

$$d^\mu = M^{\mu\nu}u_\nu = \left\{\left(-\vec{P}\vec{u}\right), \left(-u_0\vec{P} - [\vec{M}\vec{u}]\right)\right\}, \tag{1.29}$$

$$m_\sigma = \widetilde{M}_{\sigma\rho}u^\rho = \left\{\left(\vec{M}\vec{u}\right), \left(-u_0\vec{M} + [\vec{P}\vec{u}]\right)\right\}. \tag{1.30}$$

If we use the relations (1.29−1.30) in the representation (1.29), we obtain the matrix elements of the matrix (1.25).

We define the lagrangian of the interaction an electromagnetic fields with the media having polarizations as follows:

$$\mathcal{L}_I = \frac{1}{4} M^{\mu\nu} F_{\mu\nu} \ . \tag{1.31}$$

Substitute the matrix representation $M^{\mu\nu}$ (1.25) and $F_{\mu\nu}$ (1.4) in the definition of (1.31). The result is:

$$\mathcal{L}_I = \frac{1}{4} M^{\mu\nu} F_{\mu\nu} = \frac{1}{2}[(\vec{P}\vec{E}) + (\vec{M}\vec{B})]. \tag{1.32}$$

It follows from the relation (1.32) that $\mathcal{L}_I$ is an invariant regarding the Lorentz transformations.

The lagrangian consisting of the sum of the lagrangians an electromagnetic field and the interaction of this field with the polarization medium has the form:

$$\mathcal{L}_I = -\frac{1}{2}(\vec{B}^2 - \vec{E}^2) + \frac{1}{2}(\vec{P}\vec{E} + \vec{M}\vec{B}) = \frac{1}{2}(\vec{D}\vec{E} - \vec{B}\vec{H}), \tag{1.33}$$

where $\vec{D} = \vec{E} + \vec{P}$, $\vec{H} = \vec{B} - \vec{M}$, $\vec{D}$ – is a vector of electric induction, $\vec{H}$ – is a tension and $\vec{B}$ – a magnetic induction vector.

The interaction Hamiltonian $H_I$ is defined in terms of the polarization current and the potential of the electromagnetic field

$$H_I = \frac{1}{2} j_P^\mu A_\mu = \frac{1}{2}(\partial_\rho M^{\rho\mu}) A_\mu. \tag{1.34}$$

Using the relation

$$\partial_\rho(M^{\rho\mu} A_\mu) = (\partial_\rho M^{\rho\mu}) A_\mu + M^{\rho\mu}(\partial_\rho A_\mu)$$

and by the asymptotic condition, we get

$$H_I = -\frac{1}{4} M^{\rho\mu} F_{\rho\mu} = -\frac{1}{2}[(\vec{P}\vec{E}) + (\vec{M}\vec{B})]. \tag{1.35}$$

If we use the relations (1.12) and (1.28), then the lagrangian (1.32) will take the form:

$$\mathcal{L} = \frac{1}{2}[d^\mu e_\mu - m^\mu b_\mu]. \tag{1.36}$$

We now establish the connection of $d^\mu$ and $m^\mu$ with the vectors $e_\mu$ and $b_\mu$ using the polarizabilities $\alpha_E$ and $\beta_M$. To do this, we will use a relativistic generalization of the material equations $\vec{D} = \varepsilon\vec{E}$ and $\vec{B} = \mu\vec{H}$ [6,7].

If we take into account the definitions $\rho_P$ (1.22) and $\vec{j}_P$ (1.23), then Maxwell's equations (1.18) and (1.19) in the relativistic representation have the form:

$$\partial_\mu H^{\mu\nu} = j^\nu, \tag{1.37}$$

where $H^{\mu\nu} = F^{\mu\nu} - M^{\mu\nu}$.

The four-dimensional generalization of the relation $\vec{D} = \varepsilon\vec{E}$ is determined by the equation [7]

$$H^{\mu\nu}u_\nu = \varepsilon F^{\mu\nu}u_\nu. \qquad (1.38)$$

Using the vectors $d^\mu$ and $e^\mu$, the equation (1.38) can be represented as follows:

$$(e^\mu - d^\mu) = \varepsilon e^\mu. \qquad (1.39)$$

If we use the definition $\varepsilon = 1 + 4\pi\alpha_E$ then from (1.39) it follows

$$d^\mu = -4\pi\alpha_E e^\mu. \qquad (1.40)$$

When the particle velocity is equal to zero then from the relations (1.7), (1.29) and (1.40) it follows

$$\vec{\mathrm{d}} = 4\pi\alpha_E \vec{E}.$$

The relativistic generalization of the relation $\vec{B} = \mu\vec{H}$ is represented by the equation [6,7]:

$$\tilde{F}^{\mu\nu}u_\nu = \mu\tilde{H}^{\mu\nu}u_\nu, \qquad (1.41)$$

where $\tilde{H}^{\mu\nu}$ is the dual tensor of the tensor $H^{\mu\nu}$ and is expressed in terms of dual tensors $\tilde{F}^{\mu\nu}$ and $\tilde{M}^{\mu\nu}$:

$$\tilde{H}^{\mu\nu} = \tilde{F}^{\mu\nu} - \tilde{M}^{\mu\nu}. \qquad (1.42)$$

Substituting (1.42) in (1.41) and using the relation

$$\mu = 1 + 4\pi\beta_M \qquad (1.43)$$

we get the following

$$m^\mu = 4\pi\beta_M b^\mu. \qquad (1.44)$$

According to equations (1.10) and (1.30), when a microparticle velocity is equal to zero, it follows from (1.44) that the magnetic dipole moment is equal to:

$$\vec{m} = 4\pi\beta_M \vec{B}.$$

Let us now represent the lagrangian of interaction (1.31) using the definition of vectors $d^\mu$ (1.40) and $m^\mu$ (1.44). In this case the lagrangian of interaction (1.36), taking into account (1.40) and (1.44), takes the form:

$$\mathcal{L}_I = -\frac{4\pi}{2}\left[\alpha_E\left(e^\mu e_\mu\right) + \beta_M\left(b^\mu b_\mu\right)\right] = -2\pi[\alpha_E e^2 + \beta_M b^2]. \qquad (1.45)$$

If we use the definition of the vectors $e^\mu$ and $d^\mu$ through the tensors of the electromagnetic field (1.7), it is not difficult to make sure that

$$\mathcal{L}_I = -2\pi(\alpha_E F^{\mu\rho}F_{\mu\nu} + \beta_M \tilde{F}^{\mu\rho}\tilde{F}_{\mu\nu})u_\rho u^\nu. \qquad (1.46)$$

In its turn, since it follows from (1.7) that

$$e^2 = e^\mu e_\mu = \left[\left(\vec{u}\vec{E}\right)^2 - (u^0)^2\vec{E}^2 + 2u^0\left(\vec{u}\left[\vec{E}\vec{B}\right]\right) - \vec{u}^2\vec{B}^2 + \left(\vec{u}\vec{B}\right)^2\right], \qquad (1.47)$$

$$b^2 = b^\mu b_\mu = \left[\left(\vec{u}\vec{B}\right)^2 - (u^0)^2\vec{B}^2 + 2u^0\left(\vec{u}\left[\vec{E}\vec{B}\right]\right) - \vec{u}^2\vec{E}^2 + (\vec{u}\vec{E})^2\right], \qquad (1.48)$$

then the lagrangian (1.45) is expressed in terms of polarizabilities as follows

$$\mathcal{L}_I = -2\pi\{(\alpha_E + \beta_M)\left[\left(\vec{u}\vec{E}\right)^2 + \left(\vec{u}\vec{B}\right)^2 + 2u^0\left(\vec{u}\left[\vec{E}\vec{B}\right]\right)\right] - (\alpha_E(u^0)^2 + \beta_M\vec{u}^2)\vec{E}^2 -$$

$$-(\alpha_E\vec{u}^2 + (u^0)^2\beta_M)\vec{B}^2\} . \qquad (1.49)$$

It follows from the equation (1.49) that in the zero order with respect to velocity

$$\mathcal{L}_I^0 = 2\pi\left(\alpha_E\vec{E}^2 + \beta_M\vec{B}^2\right),$$

and in the expansion (1.49) to the first order in terms of velocity, we get

$$\mathcal{L}_I^{(1)} = 2\pi\left[\alpha_E\vec{E}^2 + \beta_M\vec{B}^2 - (\alpha_E + \beta_M)2\left(\vec{u}\left[\vec{E}\vec{B}\right]\right)\right]. \qquad (1.50)$$

This relationship is consistent with the lagrangian $\mathcal{L}_I$ given in [3], if we take into account the polarizabilities of the microparticle.

In summary, on the basis of Maxwell's equations and the determination of the charge density and polarization current of a structural microparticle, a relativistic tensor of the second rank is determined, the components of which are the vectors of electric $\vec{P}$ and magnetic $\vec{M}$ polarization of the medium. Using this tensor and the electromagnetic field tensor, the relativistic lagrangian of the interaction of an electromagnetic field with a structural microparticle is obtained, taking into account the polarization of its structural elements. Using the material equations in the relativistic form, the lagrangian of the interaction of an electromagnetic field with a structural microparticle is obtained, taking into account its electric and magnetic polarizabilities, and the consequences of this lagrangian are given.

### References

1. Hill, R. J. The NRQED lagrangian at order $1/M^4$ / R.J. Hill, G. Lee, G. Paz, M. P. Solon // Phys. Rev. D — 2013. — Vol. 87. — N 5. — P. 053017-1-13.

2. Silenko, A.J. Electric and magnetic polarizabilities of pointlike spin-1/2 particles / A.J. Silenko // Physics of Particles and Nuclei Letters. — 2014. — Vol. 11. — Issue 6. — P. 720-721.

3. Anandan, J. Classical and quantum interaction of the dipole/ J. Anandan// Phys. Rev. Lett. — 2000. — Vol. 85. — P. 1354-1357.

4.      Belousova, S.A. Covariant description of the interaction of an electromagnetic field with hadrons, taking into account spin polarizabilities /  S.A. Belousova (Lukashevich), O.M. Deryuzhkova, N.V. Maksimenko // Russian Physics Journal. — Vol. 43. —  N11. — 2000. —  P.  905-908.

5.      He, Xiao-Gang.  Relativistic dipole interaction and the topological nature for induced HMW and AC phases / Xiao-Gang He, Bruce McKellar //  Phys. Lett. A. — 2017. — Vol. 381. — № 21.  — P. 1780-1783.

6.      Moller, C.  The Theory of Relativity // C. Moller. — Clarendon Press. — 1957.— 386 p.

7.       Landau, L.D. Course of Theoretical Physics. Electrodynamics of Continuous Media. Vol. 8 // L.D. Landau, E. M. Lifshitz. —  Pergamon. — Edition: 2. — 1984. — 474 p.

8.      Holstein, Barry R. Hadron polarizabilities / Barry R. Holstein, Stefan Scherer// Annual Review of Nuclear and Particle Science. — 2014.  — Vol. 64. — P. 51-81.